# Superconducting thin films of MgB$_2$ on (001)-Si by pulsed laser deposition

A. Brinkman, D. Mijatovic, G. Rijnders, V. Leca, H.J.H. Smilde, I. Oomen,
A.A. Golubov, F. Roesthuis, S. Harkema, H. Hilgenkamp, D.H.A. Blank and H. Rogalla.

Low Temperature Division, MESA$^+$ Research Institute
and Faculty of Applied Physics
University of Twente
P.O. Box 217
7500 AE Enschede
The Netherlands

**Superconducting thin films have been prepared on Si-substrates, using pulsed laser deposition from a target composed of a mixture of Mg and MgB$_2$ powders. The films were deposited at room temperature and post-annealed at 600 ºC. The zero resistance transition temperatures were 12 K, with an onset transition temperature of 27 K. Special care has been taken to avoid oxidation of Mg in the laser plasma and deposited film, by optimizing the background pressure of Ar gas in the deposition chamber. For this the optical emission in the visible range from the plasma has been used as indicator. Preventing Mg from oxidation was found to be essential to obtain superconducting films.**

In January 2001, J. Akimitsu *et al.* [1,2] reported the discovery of superconductivity in the intermetallic compound MgB$_2$. The remarkably high critical temperature at which this material undergoes the transition to the superconducting state, $T_c \approx 40$ K, has aroused great interest and has spurred many groups to explore the properties and application-potential of this novel superconductor. The initial reports were soon confirmed, *e.g.* by Bud'ko *et al.*, [3] who described the occurrence of an isotope-effect, hinting towards a phonon-mediated pairing-mechanism. Band-structure calculations also point in this direction [4]. Various scientific and technological milestones are now being reached, such as first spectroscopic studies [5], measurements of the specific heat [6], investigations on the intergranular coupling [7], and the realization of MgB$_2$ superconducting wires, as accomplished by Canfield *et al.* [8]. For many electronic applications and further basic studies, the availability of superconducting thin films will be of great importance. Here we report the fabrication of superconducting thin films of MgB$_2$ on Si-substrates by pulsed laser deposition (PLD) using a target composed of a mixture of Mg and MgB$_2$.

Two complicating factors for the fabrication of superconducting films of MgB$_2$ are the high vapor pressures at low temperatures of the magnesium and the high sensitivity of magnesium to oxidation, requiring very low oxygen partial pressures in the deposition system. In order to compensate for a possible loss of Mg in the deposition-process we have used targets composed of MgB$_2$-powder enriched with Mg. The targets were prepared from a mixture of 50 Vol. % Mg powder (Alfa Aesar, purity 99.6 %) and 50 Vol. % MgB$_2$-powder (Alfa Aesar, purity 98.0 %). These powders were carefully mixed and uniaxially pressed in the form of a pellet. Then the pellets were sintered in a nitrogen-flow for 3 hours at 640 ºC and subsequently for 10 hours at 500 ºC. It was observed that during the sintering process at 640 ºC some Mg evaporated from the pellet. For this reason we removed a surface-layer of the pellet of about 1-2 mm by grinding. It is estimated that the Mg:B ratio in the final target is considerably larger than 2

Initially, using these targets as well as using targets prepared from $MgB_2$ powders only, we did not succeed in fabricating superconducting films. Numerous attempts were made, varying the deposition temperature from room temperature to ~ 600 ºC and using various post-anneal procedures. Also depositing multi-layer stacks from elemental boron and magnesium-targets by PLD and subsequent post-annealing did not lead to superconducting films, although in few cases the resistance versus temperature curves displayed small dips in the temperature range of 30 K-40 K.

While doing these experiments it was noticed that the plasma-plume invoked by the laser-ablation mostly showed a green color, which is also typically observed in the growth of MgO. For the growth of elemental Mg a blue plasma is rather to be expected. This suggested that the Mg was likely oxidized during the deposition process, explaining the absence of superconductivity in the films. Remarkably, it was found that by varying the background-pressure of Argon gas, the color of the plasma-plume could be altered from intense green for low pressures, to a bright blue for intermediate pressures and a mixture of green and blue for higher Ar pressures. Optimal Ar pressures to obtain the blue plasma were found to depend on the target composition and the laser-energy. Using a target prepared from $MgB_2$ powder only, the optimal Ar-pressure was 0.3 mbar with a laser energy $E = 500$ mJ, and 0.15 mbar at $E = 660$ mJ. Using a pure Mg-target, the pressure was 0.22 mbar at $E = 500$ mJ and 0.2 mbar at $E = 660$ mJ. Finally, using the target prepared from the mixture of Mg and $MgB_2$ the optimal pressure was found to be 0.17 mbar at $E = 500$ mJ. The latter values were used in the deposition of the films that showed the transition to the superconducting state.

The deposition of the films took place at room temperature. A KrF excimer-laser was used with a wavelength of 248 nm. As the laser spot-size on the target was approximately 8 mm$^2$, and the energy losses in the laser beam-path are estimated to be about 30%, the energy-density at the target was about 4 J/cm$^2$. As a substrate, (001)-oriented Si was used. Prior to the deposition the native $SiO_x$ layer was removed from the substrate surface using a 1% vol. HF-etching. Before growing the film, first a Mg-target was pre-ablated for 2 minutes at 10 Hz. This was done with the purpose to reduce the oxygen background pressure in the chamber by the gettering action of the Mg. Before the pre-ablation the chamber was filled with 0.2 mbar of Ar, using an Ar-flow rate of 20 ml/min.. Subsequently, the target composed of the Mg-$MgB_2$ mixture was pre-ablated under the same conditions, first for 2 minutes at 3 Hz and then for 1 minute at 10 Hz.

For the actual deposition, the pressure in the chamber was adjusted to the optimal value of 0.17 mbar and the Si-substrate was placed in front of the target at a distance of 4.5 cm (on-axis geometry). The films were deposited at a repetition-rate of 10 Hz for 6 minutes, yielding an approximate layer-thickness of 220 nm. After deposition, the Ar-gas was pumped out of the system, which then displayed a background-pressure of $2 \times 10^{-7}$ mbar. This is below the usual background pressure of our system of about $5 \times 10^{-7}$ mbar.

A high-temperature annealing step was needed to form the superconducting phase. The *ex-situ* annealing procedure took place in a 0.2 mbar Ar-atmosphere and consisted of a rapid increase (in 4 minutes) to $T_{ann} = 600$ ºC followed by a quick cool-down to room temperature. The total annealing procedure was kept short to avoid Mg-evaporation out of the film.

Figure 1 shows a resistance versus temperature curve for a film before and after the annealing. The graph was recorded using a four-point measurement configuration with voltage-pins placed approximately 2 mm apart. The film was not structured and its dimensions were about $10 \times 5$ mm$^2$. The bias current used in the measurement was 100 µA. Prior to annealing, the film does not show signs of superconductivity. After annealing, clearly a transition to zero resistance is observed with a zero resistance $T_c$ of 12 K. Remarkably, the room temperature resistance increased by the annealing procedure. It is noted that before annealing the film looked shiny like silver and after annealing shiny black.

From Figure 1 it appears likely that the film contains multiple phases, characterized by different temperature dependencies of the resistivity. Also, in these measurements the shunting path formed by the Si-substrate is expected to influence the measured temperature dependence.

In a second measurement on the same film (shown in Fig.2), we used ultrasonically bonded Al-wires, placed about 1 mm apart. Clearly in this measurement the large hump in the resistance that was observed in Fig. 1b at around 30K is absent. The zero-resistance critical temperature deduced from this measurement was 11 K, with an onset transition temperature of about 27 K. We expect that by optimizing the deposition conditions and the annealing procedure the critical temperature can be further enhanced.

In conclusion, we have fabricated superconducting films by pulsed laser deposition using targets prepared from a mixture of Mg and MgB$_2$ powders. An essential aspect to attain superconductivity in the films is to prevent oxidation of the Mg in the laser plasma and deposited film. This was achieved by optimizing the back-ground pressure of Ar gas in the deposition-chamber and using the color of the plasma as an indicator. The availability of superconducting films is an important step towards the realization of superconducting electronics based on this intermetallic compound, for which it is especially noteworthy that the films could be prepared on Si-substrates. Also, the films can be of great value for further basic studies on this intruiging material.

The authors thank G.J. van Hummel and C.A.J. Damen for assistance and O. Dolgov and J.G.J. van den Brink for valuable discussions. This work was supported by the Dutch Foundation for Research on Matter (FOM) and the Royal Dutch Academy of Arts and Sciences.

Figure captions:

Fig 1: Resistance versus temperature characteristics for an $MgB_2$ film (a) as grown and (b) after annealing at 600º C.

Fig 2: Resistance versus temperature characteristic measured at a different position on the annealed sample of fig. 1b. The inset shows the transition to zero resistance on an enlarged scale.

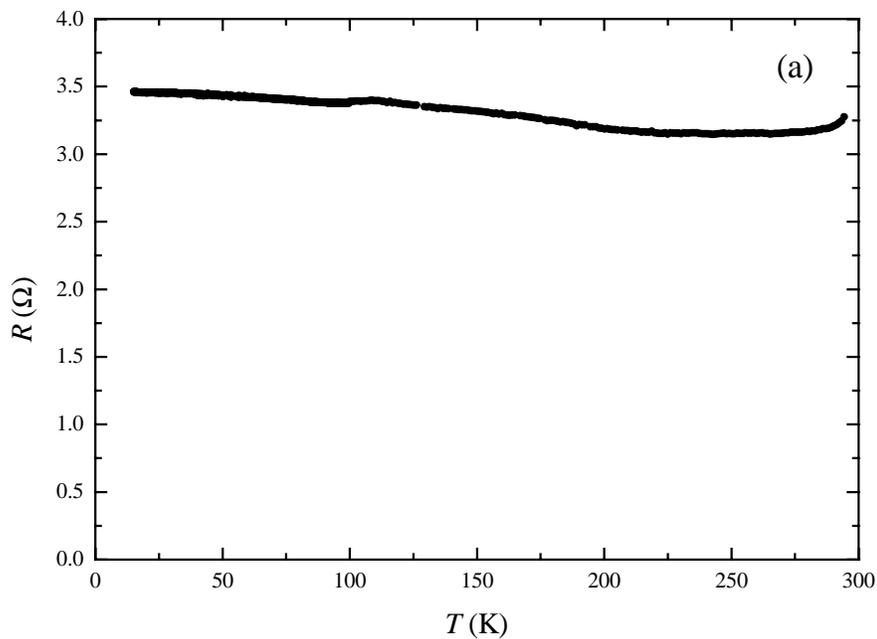

Figure 1a

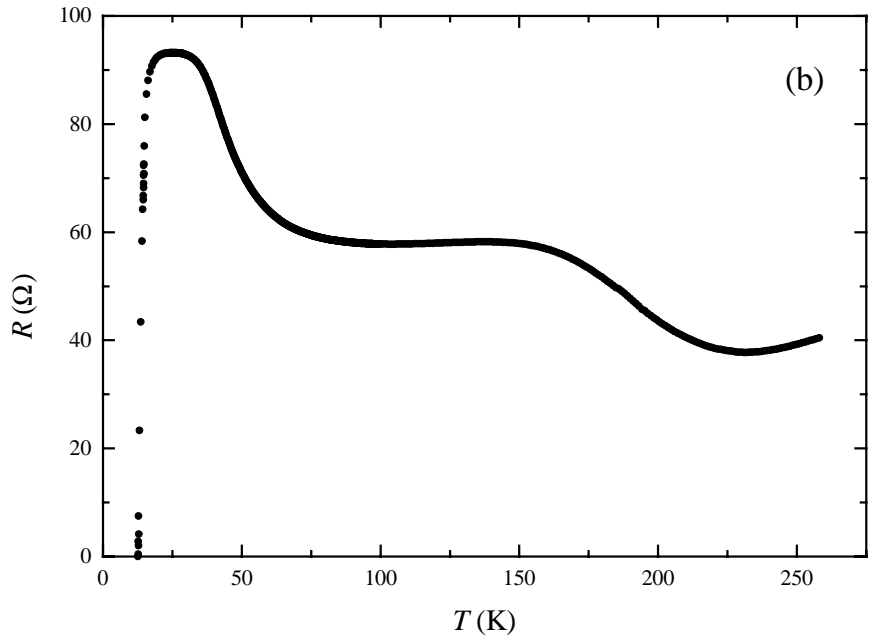

Figure 1b

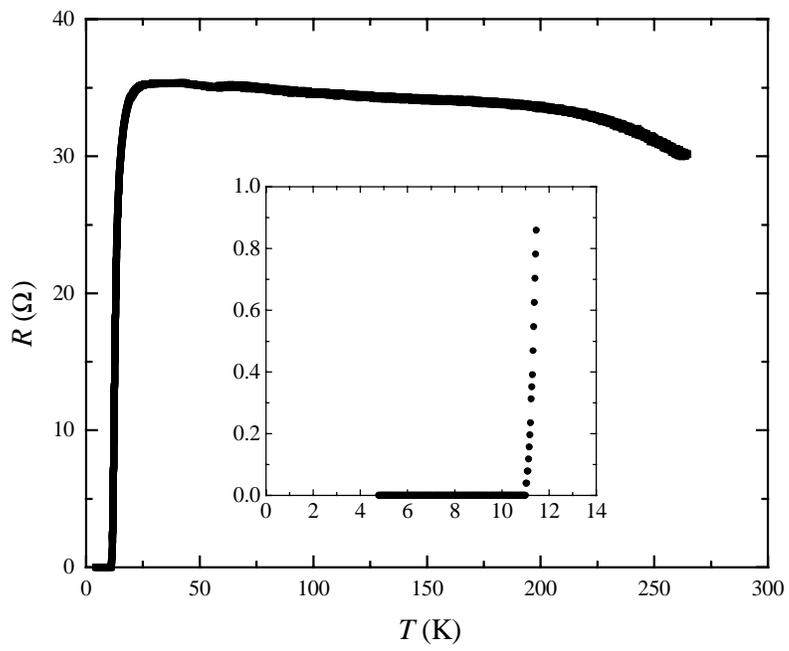

Figure 2